\documentclass[prd,aps,showpacs,nofootinbib,amssymb]{revtex4}
\usepackage{color}
\input{colordvi.tex}
\usepackage{amsmath}
\usepackage[dvips]{graphicx}
\usepackage{subfigure}
\newcommand{\beq}{\begin{equation}}
\newcommand{\eeq}{\end{equation}}
\newcommand{\beqa}{\begin{eqnarray}}
\newcommand{\eeqa}{\end{eqnarray}}

\begin{document}
\widetext
\draft

\title{Fate of thermal log type $Q$ balls}

\author{Takeshi Chiba}%
\affiliation{Department of Physics, College of Humanities and Sciences,
Nihon University, Tokyo 156-8550, Japan}
\author{Kohei Kamada}%
\affiliation{Department of Physics, Graduate School of Science,
The University of Tokyo, Tokyo 113-0033, Japan}
\affiliation{Research Center for the Early Universe (RESCEU),
Graduate School of Science, The University of Tokyo, Tokyo 113-0033, Japan}
\author{Shinta Kasuya}%
\affiliation{Department of Information Sciences, Kanagawa University, Kanagawa 259-1293, Japan}
\author{Masahide Yamaguchi}%
\affiliation{Department of Physics, Tokyo Institute of Technology, Tokyo 152-8551, Japan}

\date{\today}


\begin{abstract}
We study time evolution of the $Q$ ball in thermal logarithmic
potential using lattice simulations. As the temperature decreases due to
the cosmic expansion, the thermal logarithmic term in the potential is
eventually overcome by a mass term, and we confirm that the $Q$ ball
transforms from the thick-wall type to the thin-wall type for a positive
coefficient of radiative corrections to the mass term, as recently
suggested.  Moreover, we find that the $Q$ ball  finally ``melts down'' 
when the $Q$-ball solution disappears. 
We also discuss the effects of this phenomenon on the
detectability of gravitational waves from the $Q$-ball formation.
\end{abstract}

\maketitle

\section{Introduction}

A $Q$ ball \cite{Coleman} is a nontopological soliton, which consists of
scalar fields that carry global $U(1)$ charge $Q$. Its existence and
stability are guaranteed by finite $Q$.  $Q$ balls are interesting
objects in cosmology because they are often generated in the
Affleck-Dine (AD) mechanism for baryogenesis and can play an important
role for baryogenesis and be a good dark matter candidate
\cite{Kusenko97,Enqvist98,kk99,kk00a,kk00b,kk01,Kawasaki:2002hq}.
Recently, it was claimed that gravitational waves (GWs) are generated at
the $Q$-ball formation \cite{Kusenko:2008zm}, which may be detected by
the next generation gravitational wave detectors such as DECIGO
\cite{DECIGO} and BBO \cite{BBO}.  However, the detailed study of the
subsequent evolution and the decay of Q balls revealed it to be
difficult even by those next generation gravitational wave detectors
\cite{Chiba:2009zu}.

The properties of the AD mechanism and the $Q$ ball depend on the
supersymmetry (SUSY) breaking mechanism \cite{Kusenko97,Enqvist98},
because the effective potential of the relevant scalar field (AD field)
quite differs for the different mediation mechanism. 
Consequently, there are various types of $Q$ balls, such as gauge-mediation type
\cite{Kusenko97,kk99}, gravity-mediation type \cite{Enqvist98,kk00a},
new type \cite{kk00b}, delayed type \cite{kk01,Kawasaki:2002hq}, and so
on. Among them, the thermal log type $Q$ ball \cite{kk01}, whose effective
potential is dominated by the thermal logarithmic term \cite{Anisimov}, 
possesses an interesting feature.  As the Universe expands, the cosmic
temperature decreases and so does the thermal logarithmic potential.
Thus the properties of thermal log type $Q$ ball will change with time.
Moreover, zero-temperature potential eventually overcomes the thermal
potential and then the type of $Q$ ball changes \cite{Chiba:2009zu}. If
the zero-temperature potential itself allows a $Q$-ball solution, the type of $Q$
ball changes to the corresponding type.

It may then be naively expected that the $Q$ balls would be destroyed if the
zero-temperature potential alone does not allow a $Q$-ball
solution. Recently, however, one of the present authors (S.K.) showed that even if
the zero-temperature potential alone does not allow a $Q$-ball solution,
the total potential (the thermal logarithmic term and a mass term with a
positive radiative correction) {\it does} allow a $Q$-ball solution,
which would result in the transformation from the thermal log type of
the $Q$ ball into the thin-wall type \cite{Kasuya:2010vq}. Since the
scenario would be changed in this case, it is important to investigate
whether the field configuration dynamically transforms from one type of the
$Q$ ball to the other.

In this article we perform numerical simulations on the lattice to see
the time evolution of the configuration of the AD field in the potential
with a thermal logarithmic term and a mass term with a positive
coefficient for radiative corrections, where the latter term alone does
not allow a $Q$-ball solution.  We confirm that the thermal log type $Q$
ball transforms to the thin-wall type $Q$ ball found in Ref.
\cite{Kasuya:2010vq}.  
In addition, we find that the $Q$-ball configuration 
``melts down'' when the 
cosmic temperature becomes too low to hold 
a $Q$-ball solution. We also find that it changes the thermal history
of the Universe, but the detectability of the GWs from the $Q$-ball
formation is not improved compared to the estimate of our previous study
of Ref.~\cite{Chiba:2009zu}.

The paper is organized as follows. In the following section, we review
the properties of the thermal log type $Q$ ball and of the $Q$-ball
solution found in Ref. \cite{Kasuya:2010vq}, which we will call the thermal thin-wall type.
In Sec. \ref{sec:num}, we
show the numerical results of the time evolution of the field
configuration and confirm the transformation of the $Q$-ball types.  
In Sec. \ref{sec:melt},  we see numerically the ``melting 
down'' of the $Q$ ball as the temperature decreases further.
In Sec. \ref{sec:gw}, we reconsider the fate of the $Q$ ball and the AD
field in this case.  We also discuss the effect of this feature on the
detectability of the GWs from the $Q$-ball formation. Finally,
Sec. \ref{sec:con} is devoted to the conclusion.

\section{Properties of $Q$ balls \label{sec:thq}}

We are interested in the $Q$-ball properties in the potential with both
a thermal logarithmic term and a mass term with a positive coefficient
for one-loop radiative correction,
\begin{eqnarray}
V_{\rm tot} &=  V_{\rm thermal} + V_{\rm grav},\\ 
\label{Vth}
 V_{\rm thermal}  \simeq &\left\{
\begin{array}{ll}
T^2 |\Phi|^2, &\text{for} \ \ |\Phi| \ll T \\
T^4 \log\left(\frac{|\Phi|^2}{T^2}\right), &\text{for} \ \ |\Phi| \gg T
\end{array} \right.\label{thpot}
 \\
 V_{\rm grav} =& m_\phi^2 |\Phi|^2 \left[1+K \log \left(\frac{|\Phi|^2}{\Lambda^2}\right)\right], 
\label{gravpos}
\end{eqnarray}
where $\Phi$ is the complex AD field and $T$ is the cosmic temperature.
The upper term in $V_{\rm thermal}$ represents the thermal mass from the thermal plasma and 
the lower one represents the two-loop finite temperature effects
coming from the running of the gauge coupling $g(T)$ which depends on
the AD field value \cite{Anisimov}.\footnote{We neglect the numerical
factors coming from the gauge coupling constants.}  Note that even
before the reheating from the inflaton decay has not completed, there
exists thermal plasma from the partial inflaton decay as a subdominant
component of the Universe.  $V_{\rm grav}$ denotes a soft mass term due
to gravity-mediated SUSY breaking, where $m_\phi \sim {\cal O}({\rm
TeV})$.  The second term in the bracket is the one-loop radiative
correction, and $\Lambda$ is the renormalization scale. Here we assume
$K>0$ ($K\simeq 0.01 - 0.1$) so that $V_{\rm grav}$ alone does not allow
a $Q$-ball solution.

At larger temperature when the field starts the oscillation, the
potential is dominated by the thermal logarithmic term $V_{\rm
thermal}$, and the thermal log type $Q$ balls form
\cite{kk01,Kasuya:2010vq}. The properties of this type $Q$ ball are
similar to those of the gauge-mediation type $Q$ ball
\cite{kk01,Kasuya:2010vq},
\begin{align}
\phi_{0}(T) &\sim T Q^{1/4}, &
\omega(T)& \sim  \sqrt{2} \pi T Q^{-1/4}, & 
E(T) & \sim \frac{4 \pi \sqrt{2}}{3}T Q^{3/4}, & 
R(T) &\sim \frac{Q^{1/4}}{\sqrt{2} T} , 
\label{prop_th}
\end{align}
where $Q$ is the charge stored in a $Q$ ball,
$\phi_{0}=\sqrt{2}|\Phi_0|$ is the AD field value at the center of $Q$
ball, $\omega$ is the angular velocity of the AD field, $E$ is the
energy stored in a $Q$ ball, and $R$ is its radius.  Since the charge
$Q$ is the conserved quantity, whose value is determined at the $Q$-ball
formation, the parameters of $Q$ balls change as the temperature
decreases according to Eq. (\ref{prop_th}).  Notice that the
configuration of the AD field will follow the $Q$-ball solution, since
the time scale of the $Q$-ball reconfiguration is much shorter than the
cosmic time: $T^{-1} \ll H^{-1}$.

As the temperature decreases further, $V_{\rm grav}$ will eventually
dominate the potential at $\phi_0$. In our previous study in
Ref.~\cite{Chiba:2009zu}, we assumed that $Q$ balls are destroyed and
turn into almost homogeneous AD field quickly at this moment, because
the potential \eqref{gravpos} alone does not allow a $Q$-ball solution.
Recently, however, one of the present authors pointed out that a
$Q$-ball solution {\it does} exist even in this situation
\cite{Kasuya:2010vq}.  Although the soft mass term overcomes the thermal
logarithmic term at large field values, the latter will dominate the
potential at smaller field values.  As a result, in the light of charge
conservation, a thin-wall type $Q$-ball solution exists.

Let us investigate the condition of the existence of this $Q$-ball solution. 
A $Q$-ball solution exists when $V/\phi^2$ has a global 
minimum at $\phi=\phi_{\rm min} {\not =} 0$ \cite{Coleman}. In this case, it is satisfied when 
\begin{equation}
\frac{d}{d\phi}\left[\frac{V_{\rm tot}}{\phi^2} \right] =0 
\ \Leftrightarrow \
m_{\phi}^2 K+\frac{2T^4}{\phi^2}-\frac{2T^4}{\phi^2}\log \left(\frac{\phi^2}{2T^2}\right)=0
\label{Q:condition}
\end{equation}
has a nonzero real solution.  In fact, a $Q$-ball solution exists for
$T\gtrsim m_\phi K^{1/2}$, even after the potential is dominated by the
soft mass term at the center of the $Q$ ball. We shall call it the thermal
thin-wall type $Q$ ball.

Now we study the properties of this type of $Q$ ball further.  A
thin-wall type $Q$-ball solution satisfies \cite{Coleman}
\begin{align}
\frac{4\pi}{3}R^3&\simeq \frac{Q}{\sqrt{2 \phi_0^2V_{\rm tot}[\phi_0]}}, \\
E&\simeq Q\sqrt{\frac{2V_{\rm tot}[\phi_0]}{\phi_0^2}}.
\end{align}
For given $Q$, $\phi_0$ is determined so that the energy of 
the $Q$ ball should be the minimum.
Thus the properties of the $Q$ ball are written as
\begin{align}
\phi_0(T) &\sim c(T/m_\phi K^{1/2}) \frac{T^2}{m_\phi K^{1/2}} , & 
\omega &\sim \alpha(T) m_\phi, & 
E&\sim \alpha(T) m_\phi Q,  &
R&\sim \left(\frac{m_\phi K Q}{c \alpha T^4}\right)^{1/3} ,
\label{kasconf}
\end{align}
where $c(T/m_\phi K^{1/2})$ and $\alpha(T)$ are slowly increasing
functions of $T$ and they are of order of unity at the temperature in which we are
interested.  For example, $c(10)\simeq 2.5, c(10^2) \simeq 3.4,
c(10^3) \simeq 4.1, c(10^4) \simeq 4.6, c(10^5)\simeq 5.1, c(10^6)\simeq
5.5, c(10^7)\simeq 6.0$ and so on.  $\alpha(T)$ is expressed as
\begin{equation}
\alpha^2 = 1+K\left(\log\left(\frac{c^2T^4}{2m^2K\Lambda^2}\right)
+\frac{1}{c^2}\log\left(\frac{c^2T^2}{2m_\phi^2 K}\right)\right),  
\end{equation}
and its temperature dependence is stronger than that of $c$. 

It is true that such a $Q$-ball solution exists but not clear that the field
configuration follows from the thermal log type $Q$ ball to the
thermal thin-wall one. 
Moreover, it is nontrivial what happens when the $Q$ ball solution 
vanishes at the temperature $T\simeq m_\phi K^{1/2}$.
In order to tell how the configuration of the AD field
evolves, we perform numerical studies on the lattices
for each case in the following two sections.

\section{Time evolution and type transformation of the $Q$ ball \label{sec:num}}

In this section we investigate the evolution of the $Q$ ball in the
potential of the thermal logarithmic term and the soft mass term with
positive radiative corrections. Since we are primarily interested in the 
transformation of a $Q$ ball, here we limit ourselves to a single $Q$ ball 
assuming the spherical symmetry of the field configuration, and 
solve the one-dimensional partial differential equations in the radial direction 
by using the staggered leapfrog method with second order accuracy both in time and in space.

In order to obtain an initial configuration of the field on the
lattices, we first solve the ordinary differential equation,
\begin{equation}
\frac{d^2 \phi}{dr^2}+\frac{2}{r}\frac{d\phi}{dr}+\left(\omega^2 \phi - \frac{dV_{\rm tot}}{d\phi}\right)=0,  
\end{equation}
with  boundary conditions, $d\phi/dr(r=0)=0$ and $\phi(r=\infty)=0$ 
by using the fourth order Runge-Kutta method. 
Instead of Eq.(\ref{Vth}), here we use the following thermal potential:  
\begin{equation}
V_{\rm thermal}=T^4 \log \left(1+\frac{\phi^2}{2T^2}\right),
\end{equation}
in order to interpolate the thermal mass term at smaller field and the
thermal logarithmic term at larger field values. This form of potential includes the two limit 
in Eq. \eqref{thpot} and connects them smoothly.  Rescaling variables
with respect to $\Lambda$ [in Eq. (\ref{gravpos})], we use the following dimensionless variables:
\begin{align}
{\hat \phi}&=\frac{\phi}{\Lambda},  &
{\hat r}&=r\Lambda ,  &
{\hat E}& = \frac{E}{\Lambda}, &
{\hat t}&=\Lambda t, &
{\hat \omega}=\frac{\omega}{\Lambda}. 
\end{align}
Since we are seeking for the initial configuration when the thermal
logarithmic term dominates the potential where the $Q$ ball is the
thermal log type, we set $T_*/\Lambda=1.0\times 10^{-2}$ as the initial
temperature and $m_\phi/\Lambda = 1.0 \times 10^{-4}$.  Other parameters
are set to be $K=0.1$ and $M_G/\Lambda=24.3$, 
where $M_G$ is the reduced Planck mass. Configurations which we
find are summarized in Table \ref{Table:1}.  They coincide with the
evaluation in Eq.~\eqref{prop_th} within a numerical factor of order of
unity.

\begin{table}[htbp]
\begin{center}
\caption{Properties of the thermal log type $Q$ balls. \label{Table:1}}
\begin{tabular}{ccccc}\hline \hline
Charge ($Q$) & Angular velocity  (${\hat \omega}$)  & Energy  (${\hat E}$) 
&Field value  (${\hat \phi}_0$) &  Radius (${\hat R}$) \\ \hline
$ 2.5 \times 10^8$ & $6.0\times 10^{-4}$ & $3.9 \times 10^5$ & 2.9 & $3.2\times 10^3$ \\ 
$1.0 \times 10^9$ & $4.4 \times 10^{-4}$ & $1.1 \times 10^6$ & 4.2 &$4.5\times 10^3$ \\
$ 3.0 \times 10^9$ & $3.5 \times 10^{-4}$ & $ 2.7\times 10^6$ & 5.5 &$ 6.0\times 10^3$ \\ \hline\hline
\end{tabular}
\end{center}
\end{table}

Now we can investigate the time evolution of the $Q$ ball 
by using the second order leapfrog method. Hereafter we
assume the radiation dominated universe.
In order to guarantee the regularity at the origin, we use the variable ${\tilde \chi} \equiv \eta
{\hat r} {\hat \phi}$.  $\eta~(\equiv 2 \sqrt{{\hat t}_* {\hat t}}/a_*)$
is the conformal time rescaled with respect to $\Lambda$, where ${\hat
t}_*\equiv [90/(\pi^2 g_*)]^{1/2}\Lambda M_G/(2T_*^2)$ and $g_* \sim
200$.  $a_*$ is the scale factor at the temperature $T=T_*$ and set to
be 1. In the radiation dominated universe, $T\propto \eta \propto
a^{-1}$. We use the time step as $d\eta=0.5$.  The grid spacing is
$d{\hat r}=1.8\times 10^5/2^{16} \sim 2.7$ and the number of grid is $2^{16}$.  
We decompose the real and imaginary parts of the scalar field as 
${\tilde \chi}e^{i{\hat \omega} {\hat t}}={\tilde \chi}_R+i{\tilde \chi}_I$.  
In terms of $\eta$ and ${\hat r}$, the evolution equation of ${\tilde \chi}_i (i=R,I)$ becomes 
\begin{equation}
\frac{\partial^2 {\tilde \chi}_i}{\partial\eta^2}-\frac{\partial^2 {\tilde \chi}_i}{\partial{\hat r}^2}
+\frac{a^2 \eta {\hat r}}{\Lambda^4} \frac{\partial V_{\rm tot}}{\partial {\tilde \chi}_i}=0. 
\label{eomfornum}
\end{equation}
The boundary conditions at the origin (${\hat r} = 0$) are set to be
\begin{equation}
{\tilde \chi}_i({\hat r}=0)=0, \quad \frac{\partial{\tilde \chi}_i}{\partial\eta}({\hat r}=0)=0 
\quad (i=R, I), 
\end{equation}
and we assign the free boundary conditions at the other end of the grid. 
We have checked the charge conservation both in an expanding and
nonexpanding universe and the energy conservation in a nonexpanding
universe and have confirmed that all hold with an accuracy of $10^{-7}$ 
throughout the calculation. 
Thus, we conclude that our numerical simulation is accurate enough for our study. 

\begin{figure}[htbp]
 \begin{center}
  \includegraphics[width=80mm]{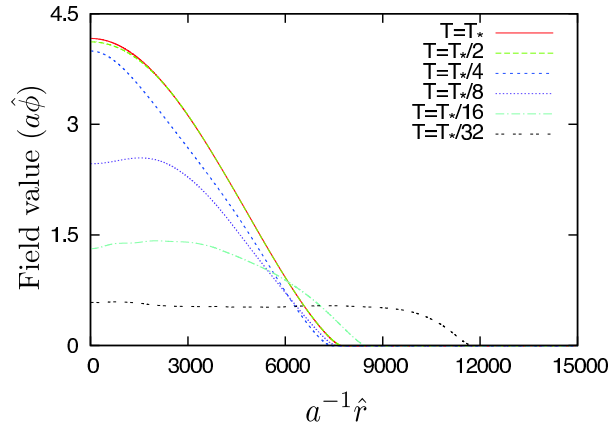}
 \end{center}
 \caption{Configurations of the AD field for $Q\simeq1.0\times 10^9$ at the time of
 $T=T_*, T_*/2, T_*/4,T_*/8,T_*/16$, and $T_*/32$ from the top to the bottom, respectively. 
 $Q$-ball configuration changes from the thick-wall to the thin-wall types.
 }
 \label{fig:confevo} 
\end{figure}

The time evolution of the field configuration is shown in
Fig. \ref{fig:confevo} for $Q\simeq 1.0\times 10^9$. 
The axes are rescaled with respect to
the scale factor $a$ so that the rescaled radius is almost constant for
the thick-wall type.  We can see that the configuration of the $Q$ ball
changes from the thick-wall to the thin-wall types.  This coincides with
the feature of the transformation of the $Q$-ball solution found in
Ref. \cite{Kasuya:2010vq}.

\begin{figure}[htbp]
 \begin{center}
  \includegraphics[width=80mm]{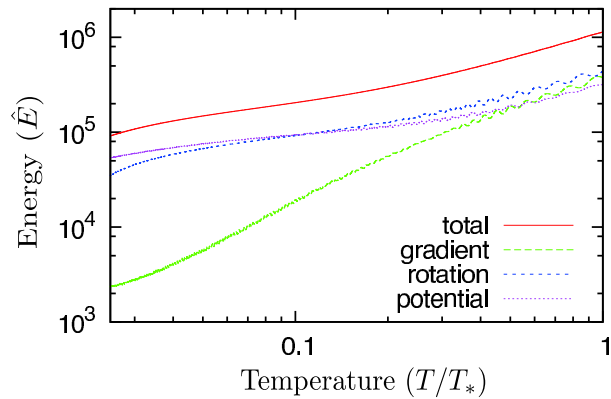}
 \end{center}
 \caption{Temperature dependence of the gradient, rotation, potential, and total energies 
 of the AD field. }
 \label{fig:enevo} 
\end{figure}

Figure \ref{fig:enevo} shows the temperature dependence or the time
evolution of the rotation, gradient, and potential energies of the field
configuration for $Q\simeq1.0\times 10^9$. The gradient energy decreases
with temperature more quickly than other contributions, and both the rotation
and the potential energy decrease in a similar manner.  These features also
imply that the thermal log type $Q$ ball becomes flattened and behaves like 
the thin-wall type one.

\begin{figure}[htbp]
\subfigure[Field value at the center of $Q$ ball]
{\includegraphics[height=5cm,keepaspectratio]{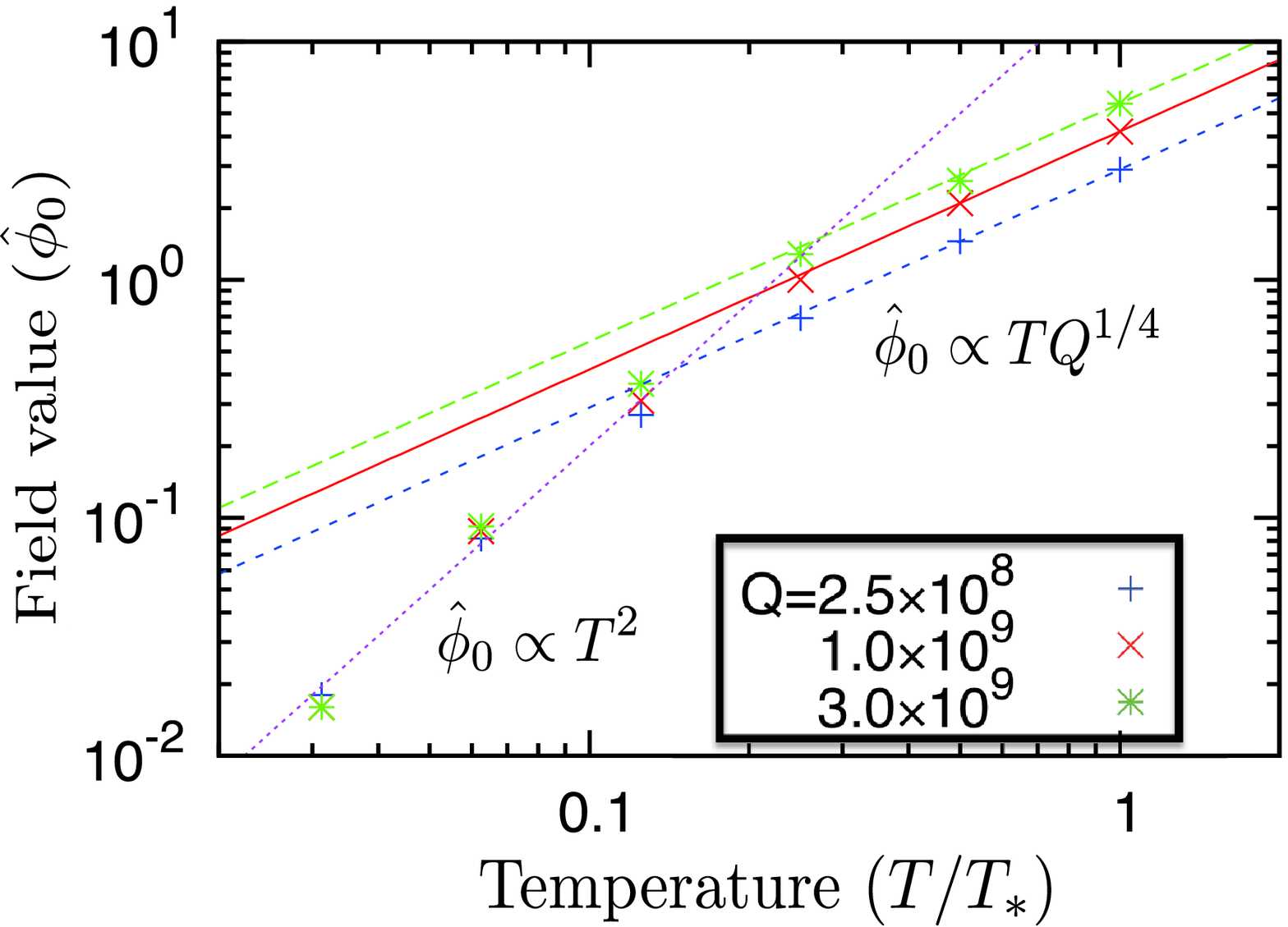}\label{RP}}
\subfigure[Radius of $Q$ ball]
{\includegraphics[height=5cm,keepaspectratio]{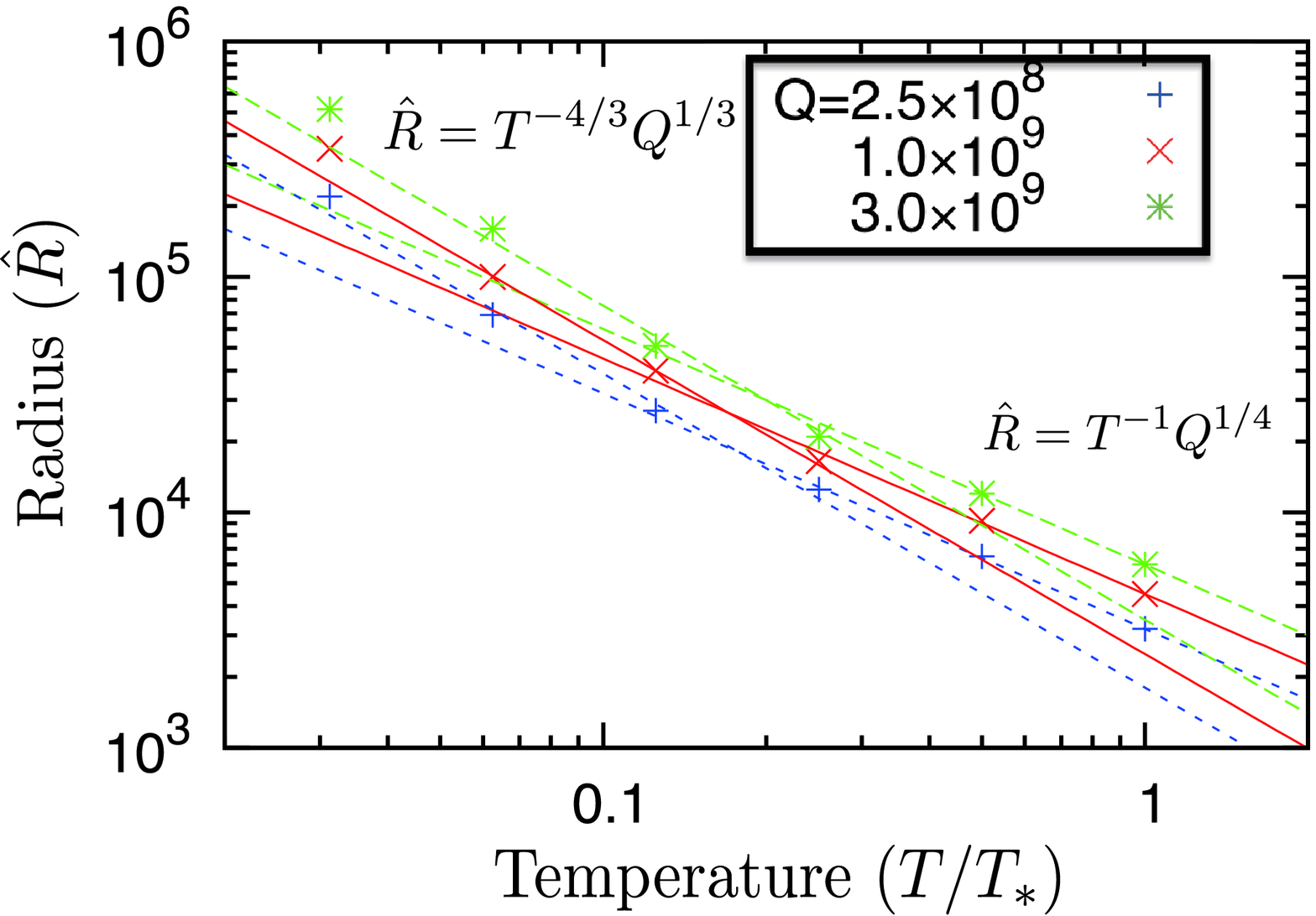}\label{RT}} 
\subfigure[Angular velocity of the AD field]
{\includegraphics[height=5cm,keepaspectratio]{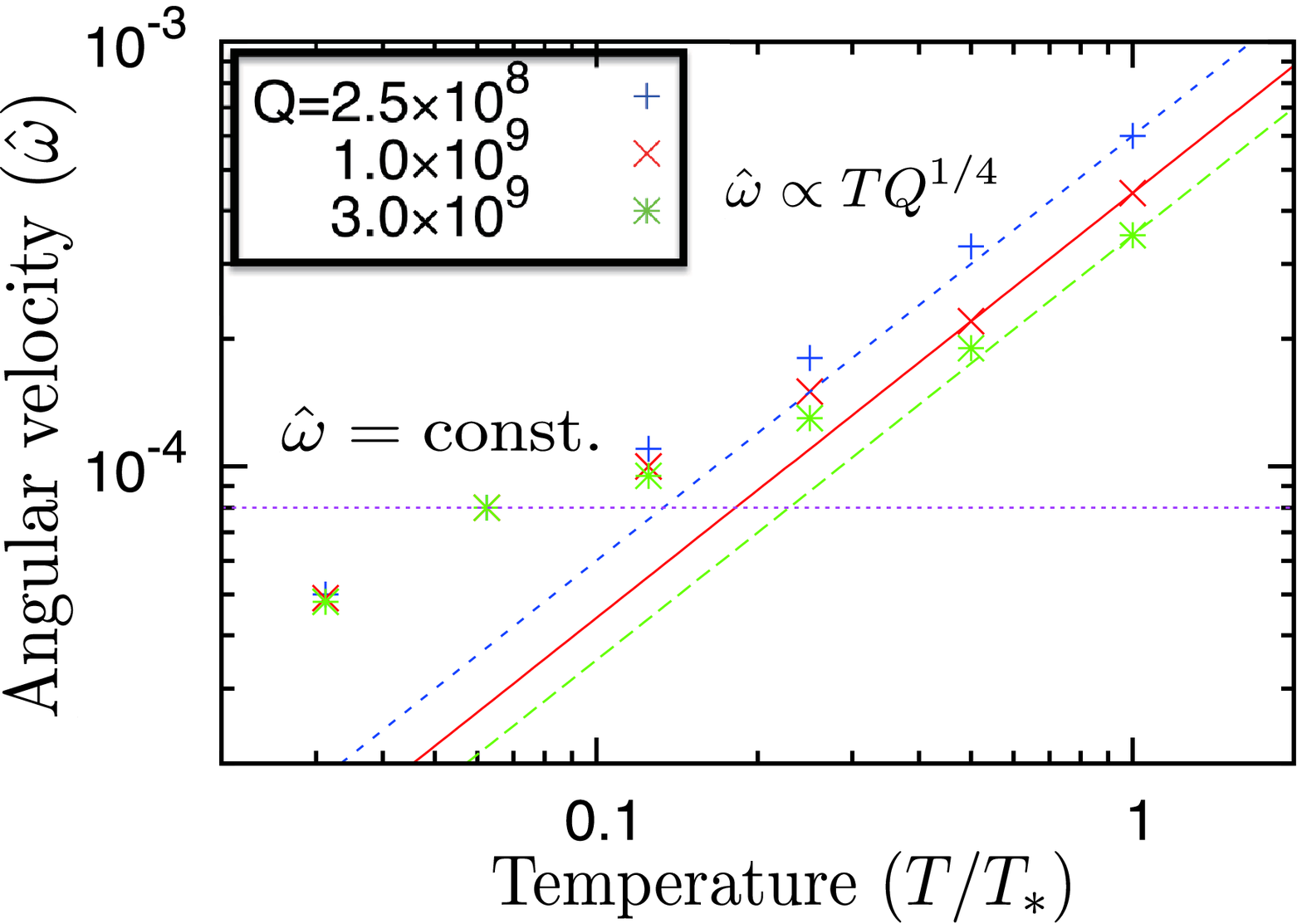}\label{Tom}} 
\caption{Temperature dependence of $Q$ ball properties. 
Crosses (blue), x's (red), and stars (green) represent the numerical results for 
$Q\simeq2.5\times 10^8, 1.0\times 10^9$, and $3.0\times 10^9$, respectively. 
Lines correspond to analytical estimates (\ref{prop_th}) and
 (\ref{kasconf}) {up to numerical coefficients}.
}
\label{fig:trp}
\end{figure}

So far we see the transformation of the $Q$-ball configuration: the
thick-wall to the thin-wall type.  We further check the properties of
the $Q$ ball before and after the transformation to confirm that they
are those of the thermal log type and the thermal thin-wall type, respectively.
The temperature dependence of the field value at the $Q$-ball center, 
the $Q$-ball radius,\footnote{
We define the radius as the length between the center of the $Q$ ball and the point 
where $\phi = \phi_0/2$.}
and the field angular velocity  with various charges 
of the $Q$ ball are shown in Fig.~\ref{fig:trp}.  
Here (blue) crosses, (red) x's, and (green)
stars represent numerical results for $Q\simeq2.5\times 10^8,
1.0\times 10^9$, and $3.0\times 10^9$, respectively. Corresponding lines 
are the analytic estimates (\ref{prop_th}) and (\ref{kasconf}) up to
numerical coefficients for each case (short dashed lines (blue): $Q\simeq2.5\times 10^8$,
straight lines (red): $Q\simeq1.0\times 10^9$, long dashed lines (green): $Q\simeq3.0\times 10^9$): At high
temperature, $\phi_0 \propto TQ^{1/4}, R \propto T^{-1}Q^{1/4},
\omega \propto TQ^{-1/4}$ for the thermal log type $Q$ ball, while, at
low temperature, $\phi_0 \propto T^2, R \propto T^{-4/3}Q^{1/3},
\omega$ = const. for the thermal thin-wall type $Q$ ball.  
We show the analytical estimates of the field value and the angular velocity in 
Eq.~(\ref{kasconf}) with
dotted lines (pueple) in Fig. \ref{RP} and \ref{Tom}, since they are independent of charge $Q$. 
We can see that the analytical estimates
 (\ref{prop_th}) and (\ref{kasconf}) are well reproduced by the
lattice simulations. One exception is the angular 
velocity at low temperature. This could be understood 
by the factor $\alpha(T)$ in Eq. \eqref{kasconf}, which decreases as $T$ gets lower. 
We can therefore conclude that the $Q$ ball really transforms from the thermal log type
to the thermal thin-wall type in the potential considered here. 

One interesting feature is the increase in the radius of the thermal thin-wall $Q$ ball 
as the temperature goes down. Since the growth rate of the radius is larger 
than that of the cosmic expansion, $Q$ balls may merge eventually.

\section{Melting of the $Q$ ball \label{sec:melt}}

In this section we investigate the evolution of the field configuration around the time  
when the potential  can no longer hold a $Q$-ball solution, 
$T\sim m_\phi K^{1/2}$ by using the numerical calculations based on the second order 
leapfrog method. 
To this end, here we set  
$T_*/\Lambda = 1.0\times 10^{-2}$, $m_\phi/\Lambda = 1.0\times 10^{-2}$, 
$M_G/\Lambda = 24.3$, and $K=0.01$ to seek the thin-wall type  
$Q$ ball as an initial configuration of the field on the lattices 
by using the fourth order Runge-Kutta method, and find the 
configurations in Table~\ref{tab:thin}. 
They coincide with the evaluation in Eq.~\eqref{kasconf} within a numerical factor of order of unity.
We then solve the equations of motion Eq.~\eqref{eomfornum} to see the time evolution of 
the field configuration on the lattices. Here we use the grid spacing  
$d{\hat r}=4.1\times 10^6/2^{19}\simeq 7.8$ and the number of the grid $2^{19}$. 
We have confirmed the charge conservation is held with an accuracy of $10^{-7}$.

\begin{table}[htbp]
\begin{center}
\caption{Properties of the thermal thin-wall type $Q$ balls. \label{tab:thin}}
\begin{tabular}{ccccc}\hline \hline
Charge ($Q$) & Angular velocity  (${\hat \omega}$) &Field value  (${\hat \phi}_0$) 
&  Radius (${\hat R}$) \\ \hline
$ 4.0 \times 10^9$ & $1.0\times 10^{-2}$ & 0.56  & $9.3\times 10^3$ \\ 
$1.1 \times 10^{10} $ & $1.0 \times 10^{-2}$ & 0.53 &$1.3\times 10^4$ \\
$ 3.0  \times 10^{10}$ & $1.0 \times 10^{-2}$ & 0.52 &$ 1.8\times 10^4$ \\ \hline \hline
\end{tabular}
\end{center}
\end{table}

\begin{figure}[htbp]
 \begin{center}
  \includegraphics[width=80mm]{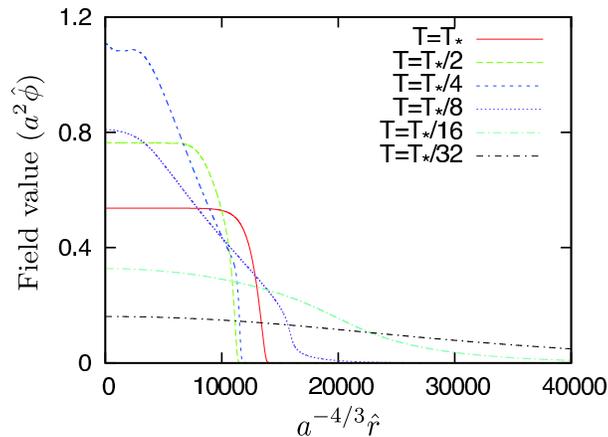}
 \end{center}
 \caption{Configurations of the AD field for $Q\simeq1.1\times 10^{10}$ at the time of
 $T=T_*, T_*/2, T_*/4,T_*/8,T_*/16$, and $T_*/32$. 
 $Q$-ball configuration starts ``melting down'' and expands.
 }
 \label{fig:melt} 
\end{figure}

\begin{figure}[htbp]
 \begin{center}
  \includegraphics[width=80mm]{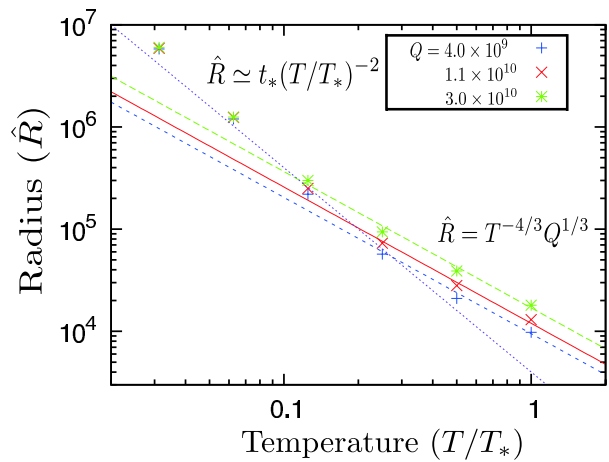}
 \end{center}
 \caption{Temperature dependence of $Q$-ball radius. Crosses (blue), x's (red) and stars (green) 
 represent the numerical results for 
 $Q\simeq 4.0\times 10^9$, $1.1 \times 10^{10}$ and $3.0\times 10^{10}$, respectively. 
 Corresponding lines are the analytical 
 estimates \eqref{kasconf} up to numerical coefficients. 
Dotted line (purple) represents the line ${\hat R} \simeq t_*(T/T_*)^{-2}$}. 
 \label{fig:meltrad} 
\end{figure}

Figure \ref{fig:melt} shows how the AD field configuration 
breaks down in the case $Q \simeq 1.1\times 10^{10}$. 
Here the vertical and horizontal axes are rescaled with 
respect to $a^2$ and  $a^{-4/3}$, respectively, 
so that the rescaled radius and field value are 
almost constant for the thermal thin-wall type. 
Temperature dependence of the radius of the $Q$ ball or AD field lump is shown in 
Fig.~\ref{fig:meltrad}, where the radius is defined as the length between the center of the 
$Q$ ball and the point where $\phi = \phi_0/10$. 
Crosses (blue), x's (red) and stars (green) represent the numerical results 
for $Q\simeq 4.0\times 10^9$, $1.1 \times 10^{10}$, and $3.0\times 10^{10}$, respectively. 
Corresponding lines are the analytic estimate 
 \eqref{kasconf} at high temperature (${\hat R} \propto  T^{-4/3} \propto a^{4/3}$) up to 
numerical coefficients for each case (short dashed line (blue): $Q\simeq 4.0\times 10^9$, 
straight line (red): $Q\simeq 1.1 \times 10^{10}$, long dashed line (green): $Q\simeq 3.0\times 10^{10}$).  
Dotted line (purple) shows the growth rate  expanding at the speed of light
$({\hat R} \simeq t_*  (T/T_*)^{-2})$ up to the numerical coefficient. 
We can see that the field configuration follows the thin-wall type of the 
$Q$-ball solution at high temperature. At the temperature $T\simeq m_\phi K^{1/2}$,
where $T \simeq 0.1 T_*$ for the parameters taken in Figs.~\ref{fig:melt} and \ref{fig:meltrad},
the $Q$-ball solution vanishes and the wall of the $Q$ ball starts 
``melting down.'' At lower temperature, 
the AD field lump expands at the speed of light and the 
configuration becomes homogeneous.  This behavior is consistent with 
the condition, Eq. (\ref{Q:condition}).


Therefore, we can conclude that the $Q$ ball starts ``melting down''
when the potential can no longer allow a $Q$-ball solution and its
remnant spreads out at the speed of light, leading to the homogeneous
AD field configuration in a few Hubble time.  Note that the $Q$ ball may
have already decayed before the ``melting'' time, since, at that time,
the field value at the center of the $Q$ ball becomes $\phi_0 \sim m_\phi
K^{1/2}$, which is as small as the electroweak scale, ${\cal O}({\rm
TeV})$.  At that time the decay channel to the fields coupled to the AD
field opens and it would decay before or during homogenization.

\section{Late time behavior of the $Q$ ball \label{sec:gw}}

In the previous sections we confirmed that $Q$ balls in the potential with
the thermal logarithmic term and the mass term that receives a positive
radiative correction transform from the thick-wall type to the thin-wall
type and then ``melt down'' into the homogeneous AD field.  In this
section, we consider the late time behavior of the time-dependent $Q$
ball and its decay time in detail.

As opposed to the thermal log type $Q$ balls whose energy density decreases as $T^4$, 
the thermal thin-wall type $Q$ balls and the homogeneous 
AD field behave like matter. Therefore, they 
would eventually dominate the Universe, 
which would alter the estimation of the detectability of the GWs from 
the $Q$-ball formation in Ref.~\cite{Chiba:2009zu}. 
In order to investigate whether they really affect the cosmic history, 
we need to estimate the lifetime of the $Q$ ball. 

There are two cases of the cosmic 
history: The $Q$ balls will decay or merge. In order to see which will 
take place, we need to estimate
 the Hubble parameter at the $Q$-ball decay, $H_{\rm dec}$,  
and the temperature at the $Q$-ball merger, $T_{\rm m}$. 
Note that the fate of the $Q$ balls differs for the different 
SUSY breaking mechanisms as discussed in Ref.~\cite{Chiba:2009zu}. 

Let us first consider the gravity-mediated SUSY breaking mechanism.
The decay rate of the $Q$ ball can be estimated as follows. 
When the field value inside the $Q$ ball is larger than the electroweak scale $E_{\rm w}$, 
the decay inside the $Q$ ball is kinematically forbidden because 
the would-be decay products acquire too a heavy mass, $g\phi_0$, for the AD field to decay into,  
and hence the $Q$ ball can decay only from its surface. Here
$g$ is the Yukawa/gauge coupling constant of order
of unity.  This sets the upper bound on the decay rate of the $Q$ ball
\cite{Cohen:1986ct},
\footnote{Originally, this bound is for the decay into fermions, which is forbidden in the 
interior of the $Q$ ball due to the Pauli principle. However, we can use this bound for the decay into scalars since 
this bound is determined by the surface region of $Q$ ball but not by the type of the decay product. }
\begin{equation}
\Gamma_{\rm dec, surf} \equiv \frac{1}{Q}\left|\frac{dQ}{dt}\right|\leq \frac{\omega^3 R^2}{48\pi Q}. 
\label{decrateq}
\end{equation}
In fact, it is known to be almost saturated for $\omega /g \phi_0\lesssim 1$ \cite{Cohen:1986ct}. 
Thus the $Q$ ball decays when the Hubble parameter becomes smaller 
than this decay rate if the field value in the $Q$ ball is large enough. 
On the other hand, when the field value inside the $Q$ ball gets smaller than the electroweak 
scale, $T\lesssim (m_\phi E_{\rm w})^{1/2}$, or $H\lesssim m_\phi E_{\rm w}/M_G$, 
the decay channel opens, $g \phi_0/\omega<1$, and 
the $Q$ ball can decay from the interior. 
At that time, the decay rate is estimated as in the case of homogeneous oscillating fields, $\Gamma \simeq h^2m/(8\pi)$, which is much larger than the Hubble parameter. 
Thus the $Q$ ball decays immediately. 
Therefore, the Hubble parameter at $Q$-ball decay is expressed as
\begin{equation}
H_{\rm dec,gr}={\rm max.} \left\{ \left(\frac{\pi^2 g_\ast}{90}\right)^{2/7}
                      \frac{\alpha K^{2/7}m_\phi^{11/7}}
                     {(48\pi)^{3/7} c^{2/7} Q^{1/7} M_G^{4/7} },\,\,
\left(\frac{\pi^2 g_\ast}{90}\right)^{1/2}
\frac{m_\phi E_{\rm w}}{M_G} \right\}. 
\label{decrate}
\end{equation}
In this estimate, we assume that the reheating temperature is high enough that
the $Q$-ball decay takes place during the radiation dominated era.

In the above argument, we have not taken into account the fact that
the thermal thin-wall type $Q$ balls would merge together and 
turn into an almost homogeneous AD field,\footnote{
The phases and the field values of the AD field in the $Q$ balls differ with each other and
hence $Q$ balls may not merge easily. However, the merger may be
completed in a few Hubble time, and hence we may assume that the merger takes place
instantaneously at $T_{\rm m,gr}$.} 
as mentioned in Sec.~\ref{sec:num}. 
This would take place if $H_{\rm m,gr}> H_{\rm dec,surf}$, where 
\begin{equation}
H_{\rm m,gr} \simeq \frac{T_{\rm m,gr}^2}{M_G}, \qquad 
T_{\rm m,gr} \simeq \frac{Km_{\phi} \phi_{\rm osc}^4}{T_R M_G^3}. \label{tmgr}
\end{equation}
Here $\phi_{\rm osc}$ is the AD field value at the onset of $Q$-ball formation and $T_R$ is the reheating temperature from the inflaton decay, and 
the distance between $Q$ balls is estimated as
\begin{equation}
L=L_{\rm f}\times \frac{a}{a_{\rm f}}=L_{\rm f}\times \left(\frac{H_{\rm f}}{H_R}\right)^{2/3}\times \frac{T_R}{T}, 
\end{equation}
where $H_R$ is the Hubble parameter at the reheating.  The distance 
between the $Q$ balls, the Hubble parameter and the $Q$-ball charge at the formation 
of the thermal log type $Q$ ball during inflaton oscillation dominated universe 
are given respectively by \cite{Chiba:2009zu,Kasuya:2010vq}
\begin{align}
L_{\rm f}&\simeq H_{\rm f}^{-1}, &
H_{\rm f}&\simeq 10^{-2} \frac{T_R^2M_G}{\phi_{\rm osc}^2},  &
Q&\simeq 2\times 10\frac{\phi_{\rm osc}^6}{M_G^2T_R^4}. 
\end{align}
The energy density of the almost homogeneous AD field is a little smaller
than the total energy density of the Universe at that time: $\rho_{\rm
AD} (\simeq m_\phi^2\phi_{\rm eq}^2\simeq T_{\rm m}^4) \lesssim \rho_{\rm tot} (\simeq (g_*
\pi^2/90)T_{\rm m}^4$), where $\phi_{\rm eq} \simeq T_{\rm m}^2/m_\phi$. 
Thus it will soon dominate the energy density of the Universe.  
Note that the thermal thin-wall type $Q$ balls will merge before they would  
dominate the energy density of the universe. 
After that, the AD field value becomes as small as the electroweak scale when 
$H\simeq m_\phi E_{\rm w}/M_G$. At that time, the AD field decay is kinematically 
allowed and takes place immediately 
since the Hubble parameter is much smaller than the decay rate. 
This is the same as the scenario that the transformation of the $Q$ ball into the 
thermal thin-wall type had not been taken account of, as in Ref.~\cite{Chiba:2009zu}. 
The only difference in the presence of the thermal thin-wall type $Q$ balls is that they 
may decay earlier from their surface before they dominate the energy density of the Universe. 

Next we consider the $Q$ ball in the gauge-mediated SUSY breaking mechanism. 
The potential is then given by
\begin{eqnarray}
\lefteqn{V_{\rm tot}=V_{\rm gauge}+V_{\rm grav2}+V_{\rm thermal}, }\\
& & V_{\rm gauge}=M_F^4\left(\log\frac{\phi^2}{2M_S^2}\right)^2,   \\
& & V_{\rm grav2}= \frac{1}{2}m_{3/2}^2 \phi^2 \left[1+K \log \left(\frac{\phi^2}{2\Lambda^2}\right)\right]. 
\end{eqnarray} 
Here  $M_S$ is the messenger mass, $M_F\simeq (m_\phi M_S)^{1/2} (> {\cal O}(1 {\rm TeV}))$,
and $m_{3/2} (< {\cal O}(10$ GeV)) is the gravitino mass, which 
is smaller than the electroweak scale. 
We are interested in the $K>0$ case where 
the thermal log type $Q$ ball transforms into 
the thermal thin-wall type  due to $V_{\rm grav 2}$.  
Once this transformation occurs, the thermal thin-wall Q balls will end up with one of the
following three destinies. The first one is that they will merge together at
the temperature 
\begin{equation}
T_{\rm m,ga} \simeq \frac{Km_{3/2} \phi_{\rm osc}^4}{T_R M_G^3}, 
\end{equation}
and turn into a homogeneous AD field rotating in $V_{\rm grav2}$. Since
the field amplitude decreases as $\phi \propto a^{-3/2} \propto T^{3/2}$
while $\phi_{\rm eq} \propto T^2$ where $V_{\rm thermal} =V_{\rm grav2}$
at $\phi=\phi_{\rm eq}$, $\phi$ never reaches to $V_{\rm thermal}$.
However, since $V_{\rm tot}$ is eventually overcome by $V_{\rm
gauge}$ at $\phi$ as the AD field value drops, the field $\phi$ gets to $V_{\rm
gauge}$ to feel instabilities and the ``delayed type'' Q balls form. They  
will decay when $H = \Gamma_{\rm dec,delay}$, where
\cite{Chiba:2009zu}\footnote{
If the mass of the decay product is larger than $\omega$, the $Q$-ball decay 
is kinematically forbidden.  In the case of the $B$ ball that carries baryonic charge, 
$\omega$ must be larger than the proton mass for its decay, which is hardly satisfied in the 
gauge-mediated SUSY breaking mechanism.  However, in the case of
$L$ ball that carries leptonic charge but not baryonic charge, $Q$ balls
can decay if $\omega$ is larger than the lightest neutrino mass.  In
addition, in the case where the $Q$ balls consist of a flat direction
with $B-L=0$, there is no such constraint.}
\begin{equation}
\Gamma_{\rm dec,delay} \simeq \frac{\pi^2}{24\sqrt{2}} \frac{m_{3/2}^5}{M_F^4}. 
\end{equation}
This is the same as the scenario which we did not consider the transformation of the 
$Q$ ball into the thermal thin-wall type in Ref.~\cite{Chiba:2009zu}.

The second is that the thermal thin-wall type $Q$ ball decays from its surface. Using 
Eq.~\eqref{decrateq}, we obtain the decay rate as
\begin{equation}
\Gamma_{\rm dec,thermal}= \left(\frac{\pi^2 g_\ast}{90}\right)^{2/7} 
\frac{\alpha K^{2/7}m_{3/2}^{11/7}}{(48\pi)^{3/7} c^{2/7} Q^{1/7} M_G^{4/7}}.
\end{equation}   

The third is that the thermal thin-wall type $Q$ ball ceases reconfiguration according 
to Eq.(\ref{kasconf}) at $T\simeq M_F$ when $V_{\rm gauge}$ starts to overcome $V_{\rm thermal}$.
Henceforth, the thin-wall type $Q$ ball has fixed configuration and decays when 
$H=\Gamma_{\rm dec,fixed}$, where
\begin{equation}
\Gamma_{\rm dec,fixed}=\frac{\alpha^{7/3} K^{2/3}m_{3/2}^{11/3}}{48\pi c^{2/3} Q^{1/3} M_F^{8/3}}.
\end{equation}   
Therefore, the transformation into the thermal thin-wall type could affect the cosmic scenario
in such a way that the decay of the $Q$ ball may take place later because the charge of the 
thermal and/or fixed thin-wall type $Q$ ball will be larger than that of the delayed type. 

Finally, we comment on the influence of the thermal thin-wall type $Q$ balls on the 
detectability of the GWs from the $Q$-ball formation. 
The present density parameter and the frequency of the GWs from the $Q$-ball formation 
can be generally written as \cite{Chiba:2009zu}
\begin{align}
\Omega_{\rm GW}^0 &= \Omega_{\rm GW}^{\rm f}\left(\frac{a_{\rm f}}{a_0}\right)^4 \left(\frac{H_f}{H_0}\right)^2, \\
f_0 &=  f_{\rm f} \left(\frac{a_{\rm f}}{a_0}\right). 
\end{align}
where the subscript ``0'' denotes that the quantity is evaluated at present.
These values would change if there is an additional $Q$-ball dominated era.
In our previous study \cite{Chiba:2009zu}, we found that when $T_R
\simeq 10^{10}$ GeV, $\phi_{\rm osc}\simeq M_G$, $M_F \simeq 10^4$ GeV,
and $m_{3/2} \simeq 10$ GeV in the gauge-mediated SUSY breaking
mechanism, GWs from the $Q$-ball formation can be detected by DECIGO or
BBO, but otherwise cannot be detected since their amplitude is too low
and/or their frequency is too large. With these parameters, $Q$-ball
merger takes place so that the cosmic history does not change. Thus, the
detectability of the GWs will be the same. What about the cases
with other parameters?  In our previous study in
Ref.~\cite{Chiba:2009zu}, we have found that the difficulties of the GW
detection come from not only the smallness of the present amplitude of
the GWs ($\Omega_{\rm GW}^0\lesssim 10^{-16}$) but also the largeness of
the typical frequency at present ($f_0 \gtrsim 10^3$ Hz).  In the case
of the gravity-mediated SUSY breaking mechanism, we see that the $Q$ ball
can decay earlier than the estimate in Ref.~\cite{Chiba:2009zu}, which
leads to a larger frequency of GWs. Thus the detectability of the GWs gets
worse.  In the case of the gauge-mediated SUSY breaking mechanism, we
find that the decay rate of the $Q$ ball may be much smaller than the
estimate in Ref.~\cite{Chiba:2009zu}, which, in turn, leads to a smaller
amplitude of GWs.  Thus the detectability of the GWs gets worse also in
this case.  Therefore, taking into account the transition of the $Q$
balls from the thermal log type to the thermal thin-wall type worsens
rather than improves the detectability of the GWs.

\section{conclusion\label{sec:con}}

In this article, we have investigated the time evolution of the $Q$
balls with both thermal logarithmic potential and the mass term with the
positive radiative corrections.  By means of one-dimensional lattice
simulations for the radial direction, assuming a spherically symmetric
profile, we have confirmed that the thermal log type $Q$ balls transform
to the thermal thin-wall type when the mass term overwhelms the thermal
logarithmic potential at the center of the $Q$ ball. In particular, we
have found that the field value at the $Q$-ball center and the angular
velocity of the AD field becomes the same value regardless of the charge
stored in a $Q$ ball.  Furthermore, we have confirmed that the $Q$-ball
configuration ``melts down'' when the cosmic temperature becomes too
small to hold any $Q$-ball solution.

Using this result, we have investigated the cosmic history in detail.
Once the $Q$ ball transforms into the thermal thin-wall type, the growth
rate of the $Q$-ball radius becomes larger than that of the cosmic
expansion.  As a result, $Q$ balls can merge before they decay and turn
into almost homogeneous AD fields.  Thus there are two scenarios in the
fate of the $Q$ balls: $Q$ balls will merge together, or decay. If
$Q$-ball merger takes place, the almost homogeneous AD field will 
dominate the energy density of the Universe soon after the merger.
In the case that the $Q$ balls
decay before the merger, we have found that, for some parameter regions,
the decay of the $Q$ balls can be earlier in the gravity-mediated
SUSY breaking mechanism and later in the gauge-mediated SUSY breaking
mechanism, compared to the case without considering the transformation into
the thermal thin-wall type, as in our previous study in Ref.~\cite{Chiba:2009zu}. 
However, the detectability of the GWs from the $Q$ balls is not
improved.

Although we have limited ourselves to the spherical symmetric system and
have studied the evolution of a single $Q$ ball, which we think will
capture most of the relevant features, it would be interesting to
investigate the merger of multiple $Q$ balls that may induce
cosmologically interesting phenomenon such as another GW emissions.
Since three-dimensional lattice simulations are required in order to see this in
detail, it will be left as a future study.

\acknowledgments 

We would like to thank Jun'ichi Yokoyama for useful comments. 
This work was partially supported by JSPS (KK) and 
the Grant-in-Aid for the Global COE Program ``Global Center of Excellence for 
Physical Sciences Frontier'' and ``The Next Generation of Physics, 
Spun from Universality and Emergence'' from the Ministry of Education, Culture,
 Sports, Science and Technology (MEXT) of Japan.
This work was also supported in part by 
Grants-in-Aid for Scientific Research from JSPS [No.\,20540280(TC) and
No.\,21740187(MY)] and in part by Nihon University.


\end{document}